\documentclass[twocolumn,
aps,prb,twoside,superscriptaddress,floatfix,tightenlines,nofootinbib]{revtex4-1}
\usepackage{dcolumn}
\usepackage{float}
\usepackage{bm}
\usepackage[T1]{fontenc}
\usepackage{amsmath}
\usepackage{graphicx} 
\usepackage{tabularx}
\usepackage{multirow}
\usepackage{graphics}
\usepackage{rotating} 
\usepackage{makecell}
\usepackage{hyperref}
\hypersetup{colorlinks=true, linkcolor=blue, citecolor=blue, urlcolor=blue}
\usepackage{tikz}
\usepackage[normalem]{ulem}
\usepackage[utf8]{inputenc}
\usepackage{booktabs}
\begin{document}
\title{Manifestation of structural Higgs and Goldstone modes in the hexagonal manganites} 
\author{Quintin N. Meier}
\email{quintin.meier@mat.ethz.ch}
\affiliation{Materials Theory, ETH Zurich, Wolfgang-Pauli-Strasse 27, 8093 Z\"urich, Switzerland}
\author{Adrien Stucky}
\affiliation{Universite de Geneve, Switzerland}
\author{Jeremie Teyssier}
\affiliation{Universite de Geneve, Switzerland}
\author{Sin\'{e}ad M. Griffin}
\affiliation{Materials Science Division, Lawrence Berkeley National Laboratory, Berkeley, CA 94720, USA}
\affiliation{Molecular Foundry, Lawrence Berkeley National Laboratory, Berkeley, CA 94720, USA}
\author{Dirk van der Marel}
\affiliation{Universite de Geneve, Switzerland}
\author{Nicola A. Spaldin}
\affiliation{Materials Theory, ETH Zurich, Wolfgang-Pauli-Strasse 27, 8093 Z\"urich, Switzerland}
\date{\today}

\begin{abstract}
Structural phase transitions described by Mexican hat potentials should in principle exhibit aspects of Higgs and Goldstone physics. Here, we investigate the relationship between the phonons that soften at such structural phase transitions and the Higgs- and Goldstone-boson analogues associated with the crystallographic Mexican hat potential. We show that, with the exception of systems containing only one atom type, the usual Higgs and Goldstone modes are represented by a combination of several phonon modes, with the lowest energy phonons of the relevant symmetry having substantial contribution. Taking the hexagonal manganites as a model system, we identify these modes using Landau theory, and predict the temperature dependence of their frequencies using parameters obtained from density functional theory. Separately, we calculate the additional temperature dependence of all phonon mode frequencies arising from thermal expansion within the quasi-harmonic approximation. We predict that Higgs-mode softening will dominate the low-frequency vibrational spectrum of InMnO$_3$ between zero kelvin and room-temperature, whereas the behavior of ErMnO$_3$ will be dominated by lattice expansion effects. We present temperature-dependent Raman scattering data that support our predictions, in particular confirming the existence of the Higgs mode in InMnO$_3$.
\end{abstract}

\maketitle
\section{Introduction}

\begin{figure}[h]
    \centering
    \includegraphics[width=\columnwidth]{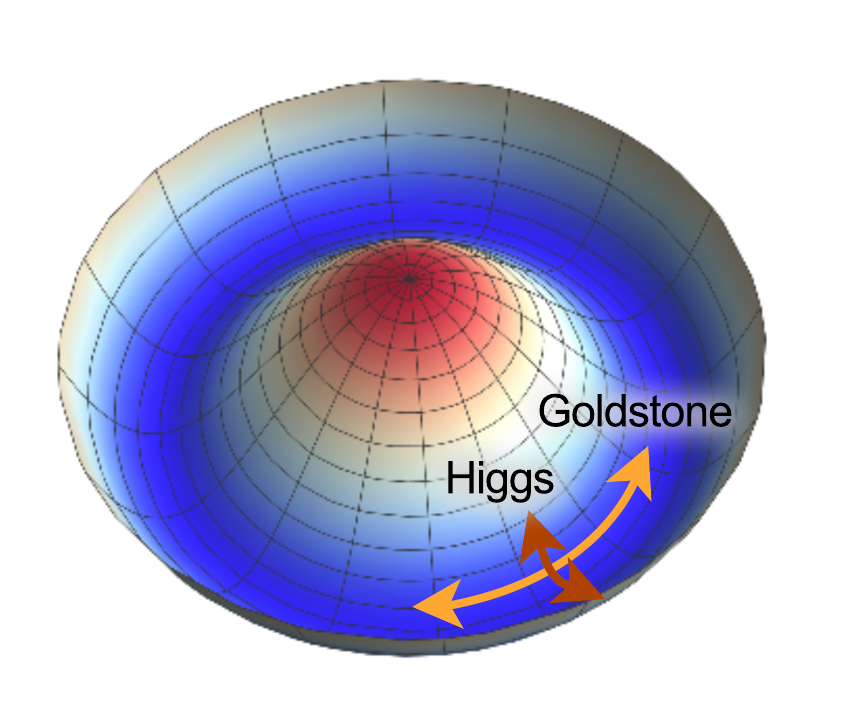}
    \caption{The Mexican hat potential describing a broken $U(1)$ symmetry, with the Higgs and Goldstone modes indicated.}
    \label{fig:U1_hat}
\end{figure}

Phase transitions that break a symmetry spontaneously occur in a wide range of physical systems, from low-energy cold atoms, through magnetic, structural and superconducting transitions in condensed matter, to high-energy collisions at the Large Hadron Collider\cite{Baumann_et_al:2010, Sadler_et_al:2006,Tinkham:1996,Chatrchyan_et_al:2012,Englert/Brout:1964,Higgs:1964,Guralnik_et_al:1964,cms, atlas}. 
Perhaps the simplest and most-studied form of spontaneous symmetry breaking is that described by the `$\phi^4$' Lagrangian, which is used in the Landau-Ginzburg theory of phase transitions, as well as in the standard model of particle physics,
\begin{equation}
\mathcal{L}=\frac{1}{2}(\partial_{\mu} \phi )^{2} - \frac{1}{2}m^2\phi^2 - \frac{\lambda}{4!}\phi^4 \quad ,
\label{L_FieldTheory}
\end{equation}
where $\phi$ is a complex order parameter which is zero in the disordered phase at $T > T_{C}$ and acquires a non-zero value below $T=T_{C}$. The energy density of such a Lagrangian has the so-called `Mexican hat' potential with continuous $U(1)$ symmetry (see Fig.~\ref{fig:U1_hat}) in which the ground-state value of the field, $\phi_{0}$, is degenerate in energy around the entire 360$^{\circ}$ rim of the Mexican hat potential. This form was originally suggested by Landau to describe ferromagnets near the critical point\cite{Landau/Lifshitz:1958, Kadanoff_et_al:1967, Peskin/Schroeder:1995} and has recently achieved notoriety following the discovery of the Higgs boson, whose formation it also describes. 

Perturbing the field $\phi$ around the ground state $\phi_{0}$ gives two types of fluctuations -- the Higgs and Goldstone modes -- which correspond respectively to oscillations of the amplitude (towards and away from the peak of the hat) and phase (around the brim of the hat) of the broken continuous symmetry (Fig.~\ref{fig:U1_hat}). 

Since the energy of the field is invariant with phase, the Goldstone mode is characteristically massless with zero frequency and a corresponding zero energy gap\cite{Goldstone:1961, Goldstone_et_al:1962}.  Many manifestations of the Goldstone mode are known in condensed-matter systems: For example a massless spin wave has been measured using neutron scattering in the prototypical Heisenberg ferromagnet EuS\cite{Boni_et_al:1995}, for which the Hamiltonian is invariant under the rotation of spins. Inelastic neutron scattering was also used to detect a gapless mode in a Bose-Einstein condensate of spin-triplet states in TlCuCl$_{3}$\cite{Ruegg_et_al:2003}, consistent with theoretical predictions\cite{Matsumoto_et_al:2002}. Polarized Raman scattering detected the development of a peak at zero frequency with divergent intensity at the structural phase transition in Cd$_2$Re$_{2}$O$_7$ pyrochlore, which has been associated with the  Goldstone phonon\cite{Kendziora_et_al:2005}; similar behavior has been predicted for Ruddeldsen-Popper-structure PbSr$_2$Ti$_2$O$_7$\cite{Nakhmanson/Naumov:2010}. Finally, Goldstone modes also appear as sound waves at the normal-to-superfluid transition in  $^{4}$He\cite{Volovik:1992}. 
In superconductors, the Goldstone mode gains mass by its interaction with an applied external field through the Anderson-Higgs-mechanism \cite{andersonCoherentExcitedStates1958,andersonRandomPhaseApproximationTheory1958}, giving rise to the Meissner effect.

In contrast, the Higgs, or amplitude, mode is massive and harmonic\cite{Higgs:1964}, with a finite energy excitation gap. It is harder to detect, because it can decay into Goldstone bosons \cite{podolskyVisibilityAmplitudeHiggs2011a}, although successful observations have been made in condensed-matter systems for which the effective theory describing the system has a relativistic form. This is the case for the superconducting phase transition in 2H-NbSe$_2$\cite{Littlewood/Varma:1981}, which provided the first experimental evidence of the Higgs mode in a condensed-matter system\cite{Sooryakumar/Klein:1980} through its unusual Raman response, which was consistent with the occurrence of an amplitude mode of the charge density wave (CDW) order parameter. Likewise, in TlCuCl$_{3}$, neutron spectroscopy measurements of the magnetic excitations revealed a Higgs mode that softened and vanished at the pressure-induced quantum phase transition from a sea of spin-singlet pairs to a long-range antiferromagnet\cite{Ruegg_et_al:2008}. Cold atoms in two-dimensional optical lattices have provided indirect measurement of the Higgs mode at the quantum phase transition between the superfluid and insulating phases, through observation of a  finite-frequency response in the superfluid phase\cite{Endres_et_al:2012}, consistent with Monte Carlo simulations\cite{Pollet_et_al:2012} and the scaling expected for a Higgs mode. Recently, the presence of structural Goldstone and Higgs modes was suggested by first-principles calculations on a strained perovskite oxide, SrMnO$_3$\cite{Marthinsen_et_al:2018}. Finally, the observation and manipulation of a Higgs mode has recently been demonstrated in a supersolid quantum gas \cite{leonardMonitoringManipulatingHiggs2017}.  A summary of experimental efforts to observe the Higgs mode in condensed matter systems can be found in Ref.~\onlinecite{Pekker/Varma:2015}.

Notably, no occurrences of the Higgs mode corresponding to {\it structural} phase transitions have been experimentally reported to date. Such an example would be convenient, since the order parameters in structural phase transitions are usually given by the  positions of the atoms, which in turn can often be measured unambiguously and remain stable for long times. Indeed, a field theoretical treatment of both Higgs- and Goldstone phonons has recently been developed and would in principle be applicable to such a transition \citep{valloneHiggsGoldstoneModes2019}. 

Here we show, that for compounds containing multiple atomic species, an unambiguous association of specific single phonons with the Higgs and Goldstone modes can not in general be made, because the different atomic masses of the species cause the eigenvectors of the force constant and dynamical matrices to differ. Nevertheless, we show that phonon modes carrying substantial Higgs and Goldstone character can be identified, and demonstrate their existence in the multiferroic hexagonal manganite family of improper ferroelectrics, which are unusual in that they have a structural phase transition whose energy landscape is described by a Mexican-hat-like potential\cite{artyukhin_landau_2013,Griffin_et_al:2012}. By combining symmetry analysis, first-principles calculations, and phenomenological modeling, we analyse the potential and dynamical energy landscapes of two representative hexagonal manganite materials, ErMnO$_{3}$ and InMnO$_{3}$. We evaluate the signatures of Higgs-Goldstone coupling in the temperature dependence of the phonon frequencies, and separate these from frequency shifts due to thermal lattice expansion. We then use temperature-dependent Raman spectroscopy to verify the predicted behavior. We find that, while the behavior of ErMnO$_3$ is dominated by lattice expansion effects up to room temperature, Higgs-mode softening can be clearly identified in the vibrational spectrum of InMnO$_3$, which provides an almost text-book manifestation of a crystallographic Higgs mode associated with a structural phase transition.

\subsection{Structural Higgs \& Goldstone modes in multi-species crystalline materials}
\label{statphon}
\label{sec:LT}

We begin by reviewing the approximations inherent in reducing the large number of structural modes associated with the many atomic displacement degrees of freedom in a solid to the effective theory $\phi^4$ theory of Eqn.~\ref{L_FieldTheory} in terms of the Higgs and Goldstone modes. Expanding the total energy of a system of atoms around their ground-state positions in the zero-temperature structure, one obtains the total energy of the system, $E$, as the sum of its kinetic and potential energies:
\begin{align}\label{eq:tot_en_0}
    E=&\dfrac{1}{2}\sum\limits_{i}m_i(\partial_t u_i)^2+\sum\limits_{i,j}\Phi_{ij}(0)u_iu_j + \sum\limits_{ijk}\Lambda_{ijk}u_iu_ju_k\\  
    & +\dfrac{1}{4}\sum\limits_{i,j,k,l}\Pi_{ijkl}u_iu_ju_ku_l - ... \quad , \nonumber 
\end{align}
where $i\in\{1x,1y,1z,2x,2y,2z,..,Nx,Ny,Nz\}$ can be a large number, leading to a large number of modes, even when periodic boundary conditions are used to constrain $N$ to the number of atoms in the unit cell. Here $\dfrac{1}{2}\sum\limits_{i}m_i (\partial_t u_{i})^2$ is the kinetic energy of the atoms,  $\Phi_{ij}(0)=\dfrac{\partial^2 E}{\partial u_i\partial u_j}$ is the harmonic force constant matrix at zero temperature ($E$ is the internal energy and $u_i,u_j$ are displacements of the $i$-th and $j$-th atoms from their positions in the zero-temperature structure), and $\Lambda_{ijk}$ and $\Pi_{ijkl}$ are the anharmonic third- and fourth-order force constant matrices. 

At low temperatures, the amplitudes of the atomic displacements are small, the average atomic positions are unchanged from the zero-temperature positions, and the phonon eigenmodes are obtained by diagonalizing the sum of the first two terms. As temperature is increased, the anharmonic couplings become relevant, leading to two effects: First, the normal modes can no longer be separated into the zero-temperature phonons and the average positions of the atoms are changed, leading to the well-established lattice expansion in conventional solids; we treat this behavior later in Section~\ref{quasi_harmonic}. Second, in systems close to a structural phase transition, the response is dominated by an additional change of atomic positions associated with the anharmonicity of a single soft-mode coordinate; we focus on this behavior here. 

To avoid calculation of the full partition function of equation \eqref{eq:tot_en_0}, the anharmonicities are renormalized into a harmonic approximation around the new atomic positions at each temperature. A renormalized force constant matrix, $\Phi_{ij}(T) =\dfrac{\partial^2 F(T)}{\partial u_i^*\partial u_j^*} $, then describes the energy cost of small atomic displacements, $u_i^*$, away from the minimum-energy atomic coordinates at temperature $T$. In this renormalized harmonic approximation the total free energy, $F(T)$, is given by
\begin{equation}
    F(T)=F_0(T)+\dfrac{1}{2}\sum\limits_i m_i (\partial_t u_i^*)^2 + \sum\limits_{ij}\Phi_{ij}(T)u_i^*u_j^* \quad ,
\end{equation}
where $F_0(T)$ is the free energy of the minimum energy structure ($u_i^*=0$) at the temperature $T$.

Next, we use the Landau theory of phase transitions to analyze the finite-temperature force constant matrix, assuming that the temperature evolution is fully captured by the evolution of the two eigenvectors ($\phi_1$ and $\phi_2$ say) that represent the two-dimensional order parameter and form the soft mode. We can then write the free energy in the usual Landau form for a broken continuous $U(1)$ symmetry\cite{hohenbergIntroductionGinzburgLandau2015} in terms of the two-component order parameter $\phi = (\phi_1,\phi_2)$ as
\begin{equation}
    F=\dfrac{a(T)}{2}(\phi_1^2+\phi_2^2)+\dfrac{b}{4}(\phi_1^2+\phi_2^2)^2 + ...
    \label{freeenergy}
\end{equation}
which has the same form as Eqn.~\ref{L_FieldTheory}.
Here
\begin{equation}
    a(T)=\begin{cases}
    >0 &\text{if } T > T_C \\
    0 &\text{if } T=T_C \\
    < 0 &\text{if } T < T_C  \quad .
    \end{cases}
\end{equation}
This approximation implies that the temperature dependence of the energy landscape is determined entirely by the anharmonicity in these two soft modes with no anharmonic coupling to other modes, and at any temperature, $\Phi_{ij}(T)$ is diagonalized by the same basis set, with the anharmonicities confined in the subspace ($\phi_1$,$\phi_2$).

Next we calculate the eigenvalues $\dfrac{\partial^2 F}{\partial \phi_i^2}$, of these two modes, above and below $T_C$.

Above $T_C$, the expectation values of $\phi_1$ and $\phi_2$ are equivalently zero, and so
\begin{align}
\left.\dfrac{\partial^2 F}{\partial \phi_1^2}\right|_{\phi_1=0,\phi_2=0} = \left.\dfrac{\partial^2 F}{\partial \phi_2^2}\right|_{\phi_1=0,\phi_2=0}=a(T) .
\end{align}

Below $T_C$, the two modes correspond to the Goldstone and Higgs modes. The Goldstone mode has an eigenvalue of zero for all temperatures below the phase transition. The Higgs mode, in contrast, softens with increasing temperature, so that its eigenvalue goes to zero at the phase transition. Formally, we obtain the solutions below $T_C$ by minimizing the free energy, $F$, (Eqn.~\ref {freeenergy}) with respect to $\phi_{1}$ and $\phi_{2}$ to extract the expectation value of the order parameter. This yields two solutions, the trivial vacuum solution with $\left<\phi_{1}\right> = \left<\phi_{2}\right> = 0$ for $-a(T)/b > 0$, and a non-trivial solution describing a degenerate circle of vacua $\phi_{1}^2 + \phi_{2}^2 = \left<\phi\right> = -a(T)/b$, which is the Mexican hat potential. Because of the $U(1)$ symmetry we can choose  $\left<\phi_{1}\right> = 0$ and $\left<\phi_{2}\right> =\left<\phi\right>$ without loss of generality, and expand around the low-symmetry vacuum ground state to obtain the excitation modes. This gives the massless Goldstone mode, corresponding to distortions along the $\phi_1$ coordinate around the brim of the hat, 

and the massive Higgs mode, corresponding to distortions along the perpendicular $\phi_2$ coordinate. 

The frequencies of the Goldstone and Higgs modes are:
\begin{align}
   \left.\dfrac{\partial^2 F}{\partial \phi_1^2}\right|_{\phi_1=\left<\phi\right>,\phi_2=0}&=0\\
    \left.\dfrac{\partial^2 F}{\partial \phi_2^2}\right|_{\phi_1=\left<\phi\right>,\phi_2=0}&=a(T)+3b\left<\phi\right>^2=-2a(T)
\end{align}

with the eigenvalues of all other modes still temperature independent. We assume therefore a strong temperature dependence of the eigenvalue of the force-constant matrix corresponding to the Higgs mode on approaching the structural phase transition, with the remaining eigenvalues being largely temperature independent. 

Finally for this section, we emphasize that, since the Mexican hat is a potential energy surface, the Higgs and Goldstone modes are the relevant eigenmodes of the force constant matrix. There is no obvious way, however, to directly measure the eigenmodes of the force constant matrix, and therefore the phonon modes, which are readily accessible via vibrational spectroscopies, are often used as proxies.  The phonon modes, however, are eigenmodes of the {\it dynamical matrix}, which is related to the force constant matrix by (see for example Ref.~\onlinecite{dove1993introduction}):
\begin{equation}
    D_{ij}=\dfrac{\Phi_{ij}}{\sqrt{M_iM_j}}\quad ,
\end{equation}
where $M_i$ is the mass of the $i$-th atom. It is clear that the force constant and dynamical matrices have different eigenvalues and eigenvectors. As a result, the Higgs and Goldstone modes associated with a crystallographic phase transition do not correspond to single phonons, except in the special case that the system contains atoms of only one mass, $M$, in which case the eigenvectors of the force constants and dynamical matrices are the same, and the phonon frequencies corresponding to the Goldstone and Higgs modes, $\omega_G$ and $\omega_H$, are given by:
\begin{align}
    \omega_G&=\sqrt{\alpha_{1}/M}\\
    \omega_H&=\sqrt{\alpha_{2}/M}  \quad ,
\end{align}
where $\alpha_1$ and $\alpha_2$ are the corresponding eigenvalues of the force constant matrix. In a general multi-component system, however, each element of the force constant matrix must be divided by the product of the square roots of the relevant masses before diagonalization to extract the phonons, and in general the static eigenvectors of the force constant matrix do not correspond to specific single phonons. For the special case of the zero-frequency Goldstone mode, the atomic masses are not relevant, and as a result there is a zero frequency phonon for each zero frequency force constant mode, and the zero frequency eigenvectors of the force-constant matrix are identical to the atomic displacements of the corresponding zero-frequency phonon mode. The static Higgs mode, however, is a linear combination of all dynamical phonon modes with the same irreducible representation, and the entire sub-space of phonon modes with the same symmetry as the Higgs mode should exhibit the strong temperature dependence that we derived above.

\subsection{Structural phase transition in the hexagonal manganites}

\begin{figure}
    \centering
    \includegraphics[width=\columnwidth]{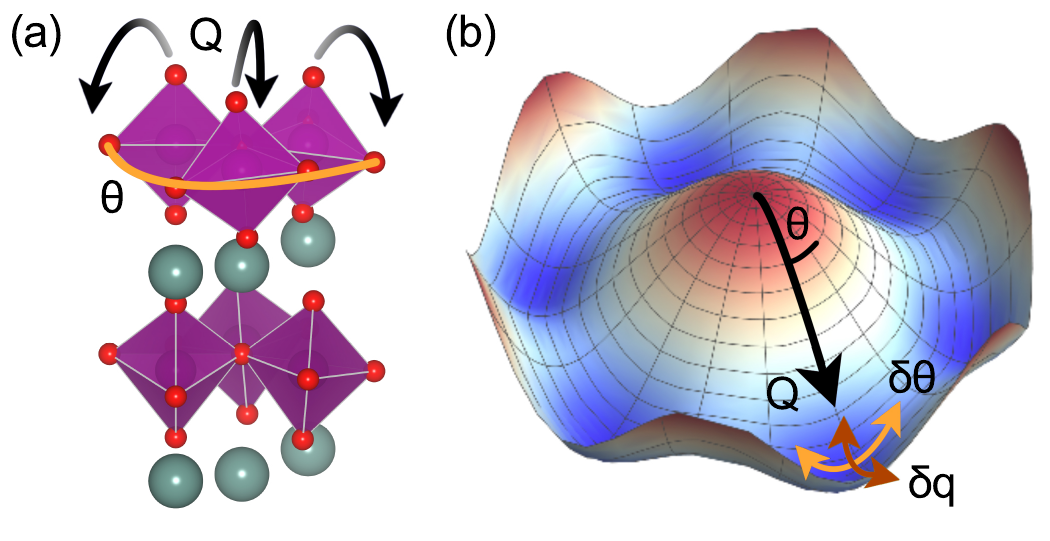}
    \caption{(a)  Crystal structure of the hexagonal manganites, blue arrows indicating the tilt amplitude $Q$ and brown circle indicating the tilt angle $\theta$ of the oxygen-manganese bipyramids. The oxygens are marked red, the manganese atoms are purple and the rare earth atoms are green. (b) Mexican-hat free energy landscape showing the energy as a function of amplitude ($Q$) and phase ($\theta$) of the order parameter. The $U(1)$-like symmetry at small order parameter values, and the six discrete minima at large order parameter values can be clearly seen. The fluctuations in the amplitude $\delta q$ (Higgs) and in the phase $\delta\theta$ (Goldstone) are indicated with red and orange arrows.}
    \label{fig:hat}
\end{figure}

The multiferroic hexagonal manganites consist of layers of corner-sharing MnO$_5$ trigonal bipyramids separating hexagonal planes of $R$ ions ($R$ = In, Sc, Y or Dy -- Lu).  They undergo a spontaneous symmetry-breaking structural phase transition between a high-temperature centrosymmetric $P6_3/mmc$ phase and a ferroelectric $P6_3cm$ structure. The primary order parameter is defined by trimerizing tilts of the MnO$_5$ trigonal bipyramids, and is two-dimensional, with its amplitude set by the magnitude of the tilt, and its phase set by the tilt angle. The crystal structure and the distortion are illustrated in Fig~\ref{fig:hat}(a). A combination of Landau theory and first-principles calculations\cite{Fennie/Rabe_YMO:2005,Artyukhin_et_al:2014,meier_2017} have shown that for small tilt amplitudes the energy is independent of the polyhedral tilt angle, and so near the phase transition the energy landscape can be described by a continuous Mexican hat potential with $U(1)$ symmetry (See Fig.~\ref{fig:hat}(b)). Thus, unusually for a crystallographic transition, the structural phase transition in the hexagonal manganites is described by a continuous primary two-dimensional order parameter ($\phi_1,\phi_2$) with an energy landscape similar to the one analyzed in the previous section. As suchit  might be expected to display Higgs and Goldstone modes. Importantly, the chemistry of the $R$ ion can strongly modify the details of the energy landscape, causing differences in the height of the peak in the Mexican hat potential and in turn influencing the transition temperature; for the ErMnO$_3$ and InMnO$_3$ considered in this work the transition temperatures are 1200 K and 500 K respectively \cite{abrahamsFerroelectricityStructureYMnO2001}.

 We note that the discreteness of the lattice manifests at larger amplitudes of the tilt mode through coupling to a secondary ferroelectric order parameter, $P$, corresponding to a net shift of the rare earth ions relative to the manganese oxygen layers along the vertical axis \cite{Fennie/Rabe_YMO:2005}. This mode has shown to be irrelevant in the region of the phase transition\cite{Jose_et_al:1977}, a concept referred to as dangerous irrelevance\cite{Amit/Peliti:1982}. The recent demonstration that the hexagonal manganites disorder continuously on all length scales close to $T_c$ \cite{skjaervoUnconventionalContinuousStructural2019} reinforces the continuous $U(1)$ behavior in the region of the phase transition.

The coupling between $P$ and the primary order parameter yields a low-temperature ground state with six minima around the brim of the hat reflecting the hexagonal symmetry so that the transition deviates from the ideal field theory of Eqn.~\ref{L_FieldTheory} and is instead described by an extended Landau free energy, which is conventionally written in the form\cite{Artyukhin_et_al:2014}: 
\begin{align}
F&=\dfrac{a}{2}Q^2+ \dfrac{b}{4}Q^4 + \dfrac{1}{6}(c + c'\cos 6 \theta)Q^6 \label{L_HexMag} \\
&-gQ^3P\cos{3\theta}+\dfrac{g'}{2}Q^2P^2+\dfrac{a_p}{2}P^2 \quad . \nonumber
\end{align}
For consistency with the hexagonal manganites literature, we use polar coordinates for the order parameter, with the amplitude $Q=\sqrt{\phi_1^2+\phi_2^2}$ and the phase $\theta=\arctan(\phi_1/\phi_2)$. 
 The energy landscape of the primary order parameter corresponds to an almost perfect Mexican hat, while the secondary order parameter induces the minima in the brim \cite{artyukhin_landau_2013}. The energy landscape for a minimized secondary order parameter is shown in figure \ref{fig:hat} (b) for the case of ErMnO$_3$. The detailed chemistry affects the coupling to the polar mode, and thus the height of the barriers around the brim of the hat. In particular in the case of InMnO$_3$ these are close to zero and the brim is  smoother than that shown here \cite{Huang_et_al:2014}. Perturbations of the phase, $\delta \theta$, and the amplitude, $\delta q$, of the order parameter are also indicated.  Perturbations of the amplitude conserve the space group symmetry, thus they belong to the irreducible representation A1. Perturbations of the order parameter angle change the space group symmetry from the ferroelectric $P6_3cm$ to $P3c1$, which corresponds to the irreducible representation B1
\cite{skjaervoUnconventionalContinuousStructural2019}.

\section{Methods}
\subsection{Sample preparation}
InMnO$_3$ samples were prepared from a stoichiometric mixture of In$_2$O$_3$ (99.9\%) and Mn$_2$O$_3$  placed in Au capsules and treated at 6 GPa in a belt-type high-pressure apparatus at 1373 K for 30 min (heating rate 110 K/min). After heat treatment, the samples were quenched to room temperature, and the pressure was slowly released. The resulting samples were black dense pellets. The energy balance between the usual polar $P6_3cm$ and an antipolar $P\bar{3}c1$ structure, formed at tilt angles half-way between those of the usual polar structures, is known to be sensitive to the details of the defect chemistry in InMnO$_3$\cite{Huang_et_al_2013,Huang_et_al:2014,griffinDefectChemistryCrystal2017}, and samples with the two phases were obtained by appropriate annealing treatment. ErMnO$_3$ samples were prepared using the PbO–PbF$_2$ flux method. The starting composition of 6.7g Er$_2$0$_3$, 7.7g MnCO$_3$, 1g B$_2$O$_3$, 7g PbO, 56g PbF$_2$ and  3.3g PbO$_2$ was heated in a 50ml Pt crucible for 15 hours at $\sim$1280$^{\circ}$C then cooled at 1$^{\circ}$C per hour to produce thin platelets around 20mm$^2$ \cite{Wanklyn:1972,Sugie/Iwata/Kohn:2002}
\subsection{Raman spectroscopy}
We performed Raman spectroscopy using a home-made spectrometer equipped with a liquid nitrogen cooled CCD camera and an Ar laser for the excitation with a wavelength of 514.5~nm. 
The Raman spectra were collected at the Stokes side of the elastic peak in the range from 50 to 850~cm$^{-1}$. Samples were mounted in a compact flow cryostat operating between 4~K and 300~K. The power of the laser was low enough to limit local heating of the sample. Of the six different irreducible representations that classify the phonon modes in the hexagonal manganites, only A1 and E2 are Raman active. 
We use two configurations of the polarization of the incoming and scattered photons: For the $z(xx)\bar{z}$ (polarization of the scattered photons is parallel to that of the incoming ones), both A1 and E2 modes are allowed by the Raman selection rules. 
For the $z(xy)\bar{z}$ configuration (scattered photons are polarized perpendicular to the incoming ones), only E2 modes can be observed. 
The relative angle of the two polarizers was calibrated using the selection rules for the 514~cm$^{-1}$ phonon line of silicon. 

\subsection{Density functional calculations}
\label{Section:DFT}
For our first-principles calculations we used density functional theory as implemented in the \textsc{abinit} code \cite{gonzeFirstprinciplesComputationMaterial2002,gonzeABINITFirstprinciplesApproach2009}. We treated the exchange-correlation functional within the LDA+$U$ approximation, with $U$ and $J$ values of 8~eV and 0.88~eV \cite{Amadon_2008}, and the core electrons using the projector augmented wave (PAW) method \cite{torrentImplementationProjectorAugmentedwave2008} from the JTH pseudopotential table provided by the abinit pseudodojo\cite{jolletGenerationProjectorAugmentedWave2014a}. We used a cutoff energy of 30~hartree and $\Gamma$-centered $k$-point meshes of $8\times8\times2$ for the 10-atom unit cells, and $6\times 6 \times 2$ for the 30-atom unit cells. Note that with these parameters InMnO$_3$ is ferroelectric; small adjustments in the parameters can stabilize the antipolar $P\bar{3}c1$ state\cite{kumagaiObservationPersistentCentrosymmetricity2012}. We obtained force constant matrices using the finite-displacement method provided in the \textsc{phonopy} package\cite{phonopy}. We calculated the Landau parameters for ErMnO$_3$ by displacing the atoms from their positions in the high-symmetry structure along the force constant eigenvectors. For our calculations within the quasi-harmonic approximation, we calculated the phonons in 30-atom unit cells and computed the internal energies and phonon free energies as a function of in-plane and out-of-plane lattice parameters. We then interpolated between the calculated values to extract the minimum energy lattice parameters at each temperature.

\section{Theoretical Results}

\subsection{Density functional calculation of zero-kelvin energetics and lattice dynamics.}

We begin by comparing the zero-temperature energetics of our two representative hexagonal manganites, ErMnO$_3$ and InMnO$_3$. As stated above, the different chemistries of the two materials lead to quantitative differences in their Mexican-hat potentials, with a hat height of $\sim$500~(170) meV and a barrier in the brim of $\sim$200~(50) meV for ErMnO$_3$ (InMnO$_3$). The  calculated Landau parameters from which these values were obtained are given in Table \ref{table:landau}. 
\begin{table}[h]
\centering
\resizebox{\columnwidth}{!}{
\begin{tabular}{c|c|c|c|c|c|c|c|}
      & $a$ [eV\AA$^{-2}$] & $b$ [eV\AA$^{-4}$] & $c$ [eV\AA$^{-6}$] & $c'$ [eV\AA$^{-6}$] & $g$ [eV\AA$^{-4}$] & $g'$ [eV\AA$^{-4}$] & $a_p$ [eV\AA$^{-2}$]\\
      \hline
      ErMnO$_3$ & -3.83 & 6.20 & 1.06 & 0.06 & 3.76 & 17.43 &  0.54  \\
      InMnO$_3$ & -0.82 & 1.13 & 0.81 & 0.04 & 1.02 & 4.85 & 3.48 
\end{tabular}{}
}
 \caption{Landau parameters for ErMnO$_3$ (calculated in this work) and InMnO$_3$ (from Ref. \onlinecite{smabratenChargedDomainWalls2018})}\label{table:landau}.
\end{table}

Next we calculate the phonon mode frequencies and eigenvectors for the two materials using density functional theory. Our calculated frequencies and the symmetries of each mode are listed in Table~\ref{Table_PhononFrequencies} of the Appendix. We see that the lowest frequency A1 mode, which we expect to have the largest Higgs character, has almost the same frequency ($\sim$130 cm$^{-1}$) in both materials, reflecting the similar curvatures of their Mexican hat potentials in the brim of the hat in the direction towards and away from the peak. The lowest frequency B1 modes, which we expect to have the strongest Goldstone character, are strikingly different however, with the frequency in ErMnO$_3$ ($\sim$107 cm$^{-1}$) considerably higher than that in InMnO$_3$ ($\sim$65 cm$^{-1}$). This is consistent with the larger barriers around the brim of the hat in the ErMnO$_3$ case. Note that even in the case of InMnO$_3$, where the brim of the hat is very smooth, the frequency is still quite far from zero.

Finally, in anticipation of differences in the Higgs-Goldstone coupling caused by the different Mexican hats, we  calculate the phonon-phonon couplings between the low frequency A1 and B1 modes in the two materials. Our results are presented in Table~\ref{appendix:phonon_coupling} of the appendix, with the form of the coupling given by
\begin{align*}
E_\text{phonon-phonon}&=\dfrac{\omega_{\text{A1}}}{2}^2A_{\omega_{\text{A1}}}^2+\dfrac{\omega_{\text{B1}}}{2}^2A_{\omega_{\text{B1}}}^2+cA_{\omega_{\text{A1}}}^3+\\
 &dA_{\omega_{\text{A1}}}A_{\omega_{\text{B1}}}^2+e A_{\omega_{\text{A1}}}^4+fA_{\omega_{\text{B1}}}^4+gA_{\omega_{\text{A1}}}^2A_{\omega_{\text{B1}}}^2
\end{align*}
For the lowest-lying A1 and B1 modes, we observe that the lowest-order coupling term, $d$, is larger in InMnO$_3$ than in ErMnO$_3$.

\subsection{Landau theory calculation of the temperature dependence of the Higgs- and Goldstone-like phonon modes.}

Next we use Landau theory based on our calculated density functional theory parameters to calculate the explicit temperature dependence of the phonon frequencies.

We begin by extending the Landau theory framework that we developed in section \ref{sec:LT} to include, in addition to the two-component primary order parameter, the secondary order parameter that is relevant in the hexagonal manganites. We then calculate explicitly the temperature dependence of the phonons in both ErMnO$_3$ and InMnO$_3$ using the coefficients of Table~\ref{table:landau}. We proceed by expanding the free energy of Eqn.~\eqref{L_HexMag} in terms of small perturbations of the primary (treating each component separately) and secondary order parameters around the minimum energy positions, $\theta=0+\delta\theta$, $Q=\tilde Q +\delta q$, $P=\tilde P+\delta p$. Here $\tilde Q$ is the expectation value of $Q$, given by the solution of the equation $\left.\dfrac{\partial F}{\partial Q}\right|_{\theta=0,P=\tilde P}$=0. Correspondingly, $\tilde P$ is the expectation value of $P$, which we obtain from the solution of $\left.\dfrac{\partial F}{\partial P}\right|_{Q=\tilde Q,\theta=0}=0$, yielding
\begin{equation}
\tilde P[Q,\theta]=\dfrac{gQ^3\cos3\theta}{g'Q^2+a_p} \quad .
\end{equation}
We obtain the following effective susceptibilities for the perturbations of each component:
\begin{align}
    \chi^{-1}_{\delta q}&=a(T) + 3b\tilde Q^2 +5b \tilde Q^4 (c+c')-6g\tilde Q\tilde P+g'\tilde P^2\label{eq:chiq}\\
    \chi^{-1}_{\delta\theta}&=-6c'\tilde Q^4 + 3g\tilde Q\tilde P\\
    \chi^{-1}_{\delta p} &= g'\tilde Q^2+a_p \quad . \label{eq:chip}
\end{align}

\begin{figure}[h]
    \centering
    \includegraphics[width=\columnwidth]{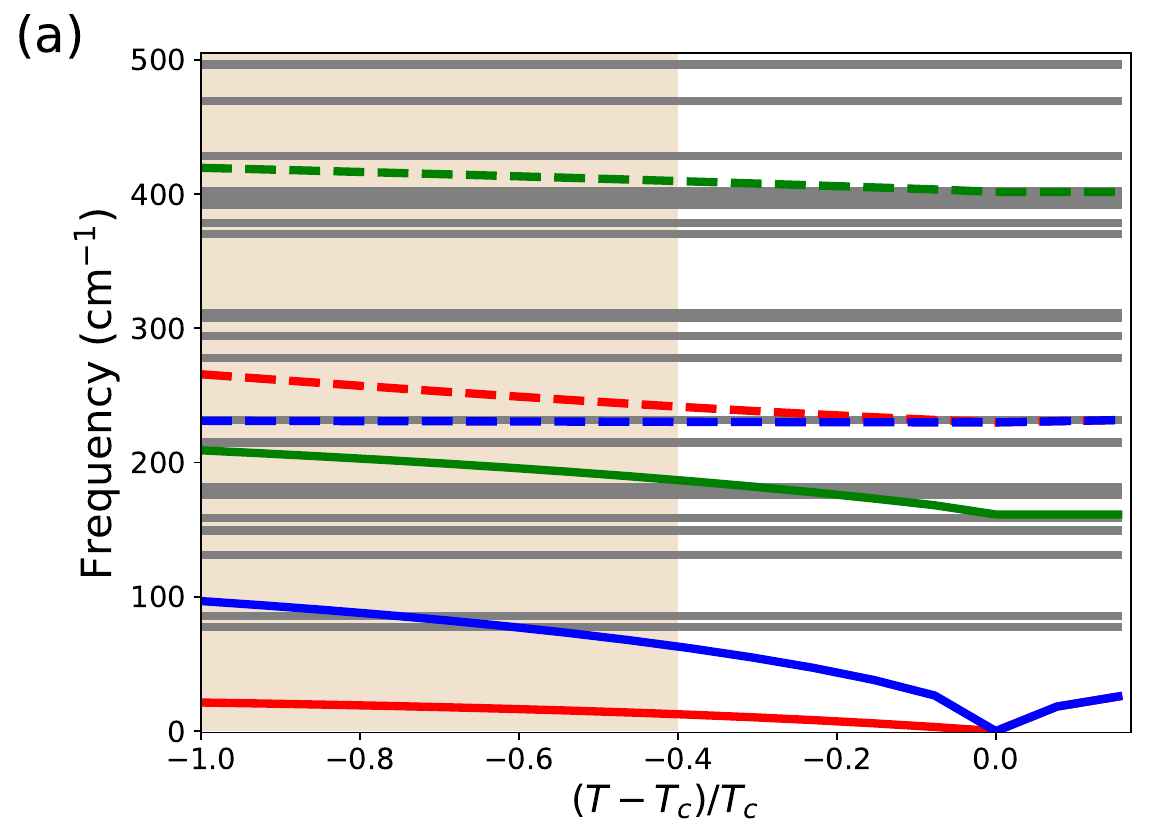}
    \includegraphics[width=\columnwidth]{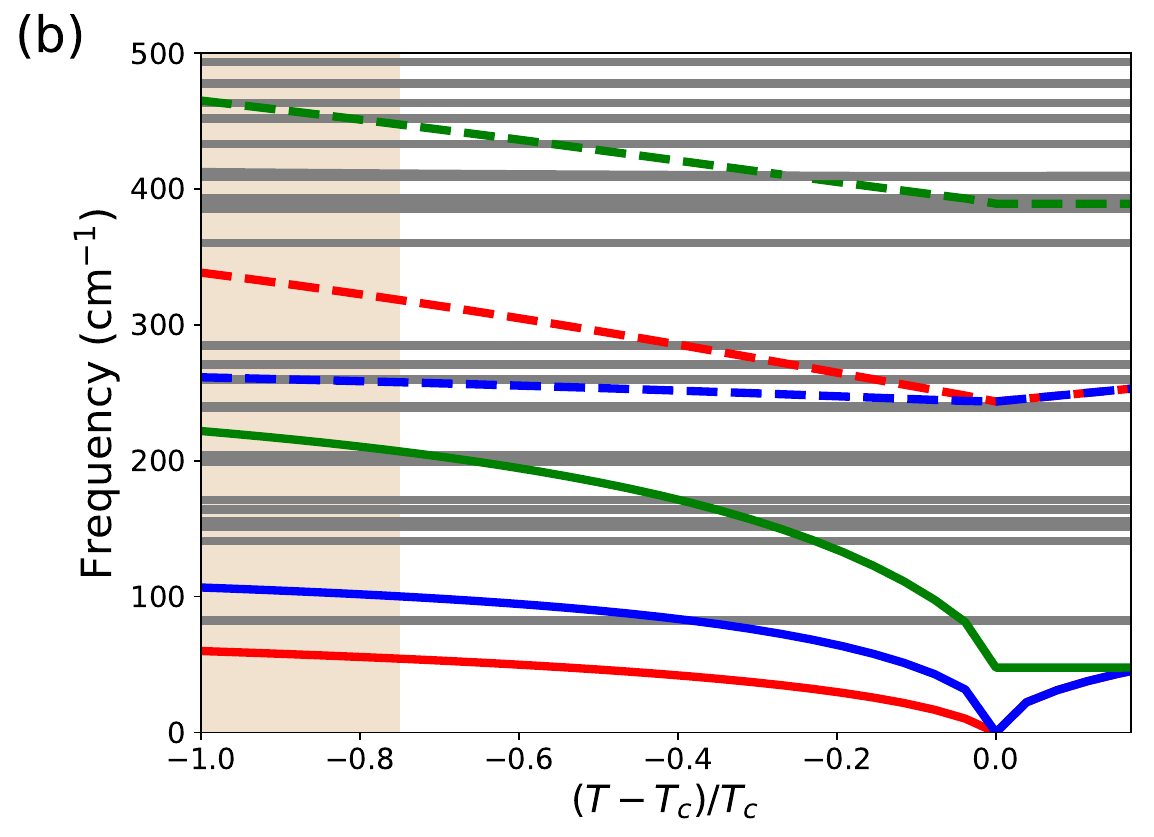}
    \caption{Calculated evolution of phonon frequencies from the Landau theory approach for (a) ferroelectric InMnO$_3$ and (b) ErMnO$_3$. Modes changing due to freeze-in of the primary order parameter are marked red (Goldstone, B1) and blue (Higgs, A1), while the phonon modes related to the polar instability (A1) are marked green. Shaded areas mark the temperature range accessible by our Raman experiments.}
    \label{fig:ev_hgs}
\end{figure}

We then calculate the phonon frequencies at each temperature by replacing the calculated zero-temperature susceptibilities by the response functions \eqref{eq:chiq}-\eqref{eq:chip} in the force constant matrix, and assuming the usual linear evolution, $a(T)=a_0(T-T_c)/T_c$, for the temperature dependence of the $a$ parameter of the soft mode. Our calculated phonon frequencies as a function of temperature are shown in Fig.~\ref{fig:ev_hgs} (a) for InMnO$_3$ and (b) for ErMnO$_3$. Modes of A1 symmetry (and therefore Higgs character) are indicated in blue, B1 symmetry (Goldstone) modes in red, and polar modes in green.  The frequencies of all the other phonons are temperature independent by construction in our approximation.

We begin by comparing, in Table~\ref{tab:freqs}, the frequencies obtained from our Landau theory approach in the zero-kelvin limit with those calculated for the fully relaxed cell using density functional theory. We find that the Landau theory frequencies underestimate the DFT values by around 30\%.  This is mostly a result of our neglecting the weak coupling of the order parameter modes to two additional modes, as described in Ref.~ \onlinecite{Fennie/Rabe_YMO:2005}. This additional coupling would harden the relevant phonon modes in the Landau description. 

\begin{table}[]
    \centering
    \begin{tabular}{l|c|c|c|c}
          & Higgs 1 & Goldstone 1 & Higgs 2 & Goldstone 2   \\
          \hline
         ErMnO$_3$ (Landau) & 106 & 82 & 285 & 338 \\
         ErMnO$_3$ (DFT) & 130 & 107 & 315 & 424 \\ \hline
         InMnO$_3$ (Landau) & 96 & 21 & 232 & 265 \\
         InMnO$_3$ (DFT) & 128 & 65 & 235 & 302 \\
    \end{tabular}
    \caption{Zero temperature frequencies of the modes with Higgs and Goldstone character in ErMnO$_3$ and InMnO$_3$ extrapolated from the Landau model and calculated using DFT for the relaxed low-symmetry structures. All frequencies are given in units of cm$^{-1}$.}
    \label{tab:freqs}
\end{table}

Next we discuss the temperature dependence of the modes, beginning with the softest mode. First, we note that this mode, which softens on approaching $T_C$ from above, is doubly degenerate above $T_C$ due to the equivalence of the order parameter directions in the high-symmetry phase. The degeneracy is lifted and the mode splits into two below the phase transition, an amplitude mode with A1 symmetry, which can be regarded as the primary Higgs mode of the structural transition (shown in blue) and a phase mode with B1 symmetry which represents the primary Goldstone mode (shown in red). The Goldstone mode in InMnO$_3$ retains a lower frequency down to zero kelvin than that in ErMnO$_3$, reflecting the smaller barriers in the brim of the hat in the InMnO$_3$ case. The Higgs modes have a similar temperature dependence and zero-kelvin frequency in both compounds. We note, however, that the zero of $(\frac{T-T_c}{T_c})$ corresponds to a very different temperature in the two cases ($\sim 1200~K$ for ErMnO$_3$ and $\sim 500~K$ for InMnO$_3$. We find, as expected, the occurrence of additional temperature-dependent Higgs-like and Goldstone-like modes at higher frequencies, indicated with dashed lines in Fig.~\ref{fig:ev_hgs}. These modes have the same symmetry as the soft modes, and their temperature dependence is a consequence of the mixing of eigenmodes caused by the transformation from the force constant to dynamical matrices. This mixing is stronger in ErMnO$_3$ than in InMnO$_3$, because of the larger mass of Er, resulting in a stronger temperature dependence of the higher frequency A1 and B1 phonons in ErMnO$_3$.

The temperature evolutions of the phonons corresponding to the polar modes are plotted in green. These are independent of temperature above $T_C$, but we find that their frequencies increase below the phase transition, as the increase in magnitude of the primary order parameter stabilizes the polar mode. We see that the frequency of the polar mode increases more in ErMnO$_3$ than InMnO$_3$, consistent with the larger coupling $g$ in the $Q^3P\cos{3\theta}$ term of the Landau free energy for ErMnO$_3$.

\subsection{Effect of change in lattice parameters on the phonon mode frequencies}
\label{quasi_harmonic}

Finally for this section, we calculate how the change in lattice parameters with temperature affects the phonon frequencies in ErMnO$_3$, with the goal of isolating any mode softening due to thermal expansion from the mode softening due to approaching the phase transition discussed above. ErMnO$_3$ and the other rare-earth hexagonal manganites are known experimentally to have an unusual lattice response to temperature, with the in-plane lattice parameter $a$ increasing with temperature as expected, but the out-of-plane $c$ lattice parameter decreasing with increasing temperature\cite{gibbsHightemperaturePhaseTransitions2011}.

We begin by demonstrating that this unusual evolution can be captured within the quasi-harmonic approximation, in which the total free energy as a function of the lattice parameters, $F_{\text{tot}}$ is obtained by minimization of the sum of the internal energy $E(a,c)$ plus the phonon free energy $F_{\text{phonons}}(a,c)$:
\begin{equation}
    F_{\text{tot}}=\min\limits_{a,c}\left[E(a,c)+F_{\text{phonons}}(a,c)\right]
\end{equation}
where $a$ and $c$ are the in-plane and out-of-plane lattice parameters. $E(a,c)$ is obtained by relaxing all internal degrees of freedom for the $P6_3cm$ structure for the set of given lattice parameters, and the phonon free energy is calculated using the partition function for harmonic phonons:
\begin{equation}
    F_\text{phonons}=\dfrac{1}{2}\sum\limits_{q\nu}\hbar\omega_{q\nu}+k_BT\sum\limits_{q\nu}\ln\left[1-\exp{\hbar\omega_{q\nu}/(k_BT)}\right]
\end{equation}

\begin{figure}[h]
    \centering
    \includegraphics[width=\columnwidth]{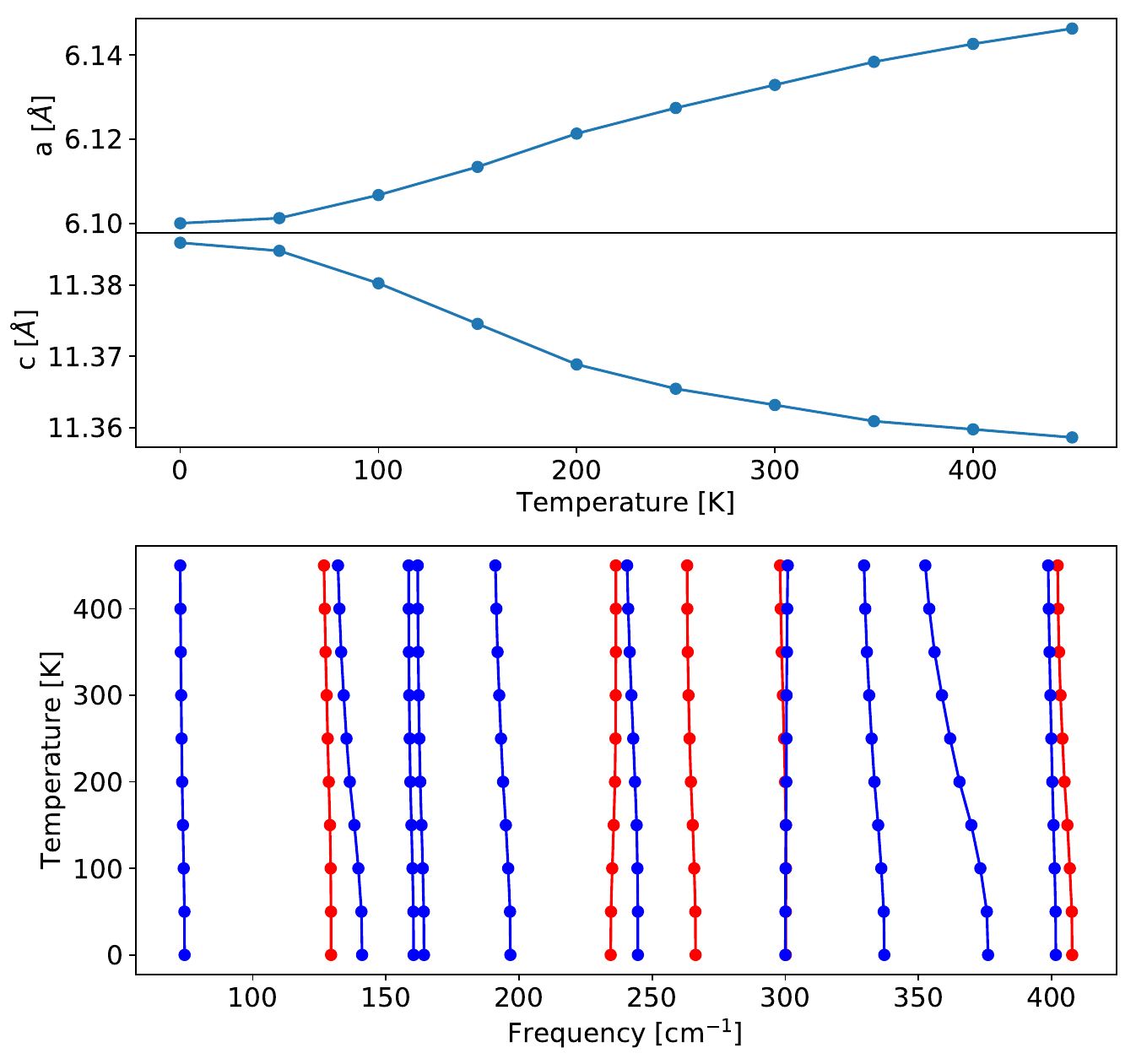}
    \caption{(a) Lattice parameters for ErMnO$_3$ calculated within the quasi-harmonic approximation. (b) Temperature dependence of the phonon frequencies calculated for ErMnO$_3$ within the quasi-harmonic approximation. We show only the A1  (red) and E2 (blue) modes for comparison with the Raman spectroscopy measurements in the next section.}
    \label{fig:qha}
\end{figure}

Using this approach, we calculate the temperature dependence of the lattice parameters, which we present in Figure \ref{fig:qha} for ErMnO$_3$.  The excellent agreement with experiment~\cite{skjaervoUnconventionalContinuousStructural2019} suggests that the quasi-harmonic population of phonons with increasing temperature is the dominant contribution to the thermal evolution of the lattice parameters. We then approximate the temperature dependence of the phonon frequencies, by calculating the eigenmodes of the dynamical matrix at the $a,c$ lattice parameters for the corresponding temperature. We deliberately omit anharmonic interactions and phonon populations in this step, in order to isolate specifically the effect of the change in lattice parameters. We show our results for the Raman-active A1 and E2 phonons in Figure \ref{fig:qha}.  We find that in this limit, most modes, in particular the A1 and B1 (not shown) phonons relevant to the Higgs-Goldstone coupling are largely temperature independent.  Therefore we can exclude that any measured temperature dependence of the Higgs and Goldstone modes is a result of the change in lattice parameters with temperature.

\section{Experimental results}
\subsection{Raman spectroscopy}
\begin{figure}[h]
\includegraphics[scale=0.25]{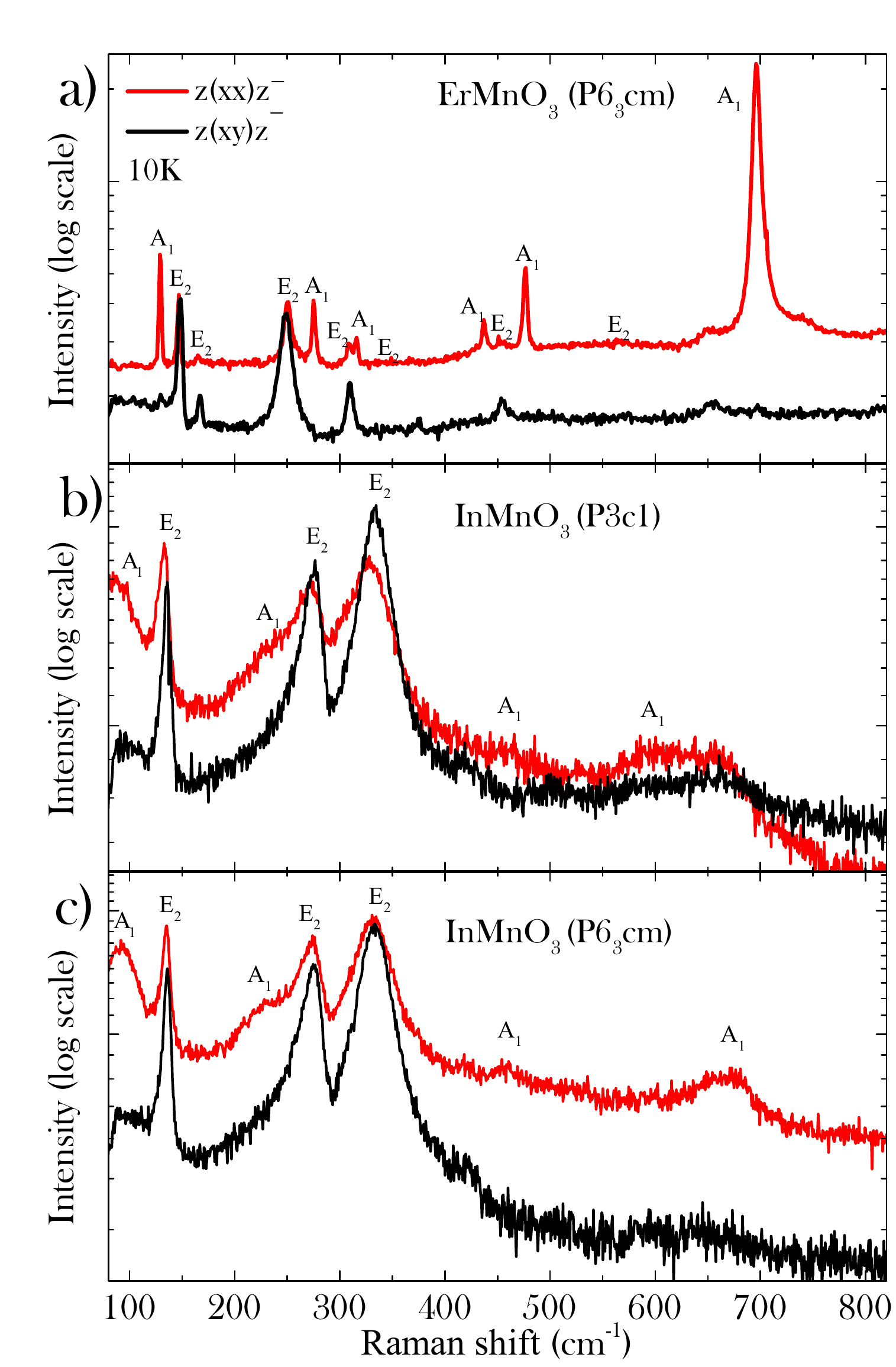}
\caption{Raman spectra collected at 10 K on single crystals of (a) ErMnO$_3$, (b) InMnO$_3$ (P$\bar{3}$c1) and (c) InMnO$_3$ (P6$_3$cm). The red and black lines show the intensity of Raman scattering in the parallel and perpendicular configurations respectively. The polarization selection rule for the ErMnO$_3$ sample ($P6_3cm$ symmetry) is clearly manifested.}\label{fig_pol}
\end{figure} 
In Fig.~\ref{fig_pol} the Raman spectra at 10~K are displayed for (a) ferroelectric $P6_3cm$ ErMnO$_3$, (b) InMnO$_3$ in the antipolar $P\bar{3}c1$ state, and (c) the ferroelectric $P6_3cm$ variant of InMnO$_3$, for parallel (red) and perpendicular (black) polarizations of incoming and scattered photons.
Our ErMnO$_3$ data (Fig.~\ref{fig_pol}(a)) are in excellent agreement with previously published results~\cite{vermetteRamanStudySpin2008}, showing all the previously reported A1 and E2 Raman active modes with the expected relative intensities and positions. The extinction of the A1 modes in the perpendicular configuration confirms the selection rules for the $P6_3cm$ space group, and the narrow linewidths confirm the high quality of the ErMnO$_3$ single crystal used in this study.
For both the $P6_3cm$ and $P\bar{3}c1$ InMnO$_3$ crystals we observe in Fig.~\ref{fig_pol}(b) and (c) the extinction of the mode at 680 cm$^{-1}$ and of the shoulder at 280 cm$^{-1}$ for the perpendicular polarizer configuration, which indicates that these modes belong to the A1 representation. The peaks at $\sim$ 140, 280 and 330 cm$^{-1}$ persist for perpendicular polarization and therefore have E2 symmetry. The small crystal size leads to broad peaks and difficulty in umambiguously assigning the remaining peak frequencies, although the peaks at around 450 and 600 cm$^{-1}$ are likely of A1 symmetry.

\begin{figure}[h]
\includegraphics[scale=0.25]{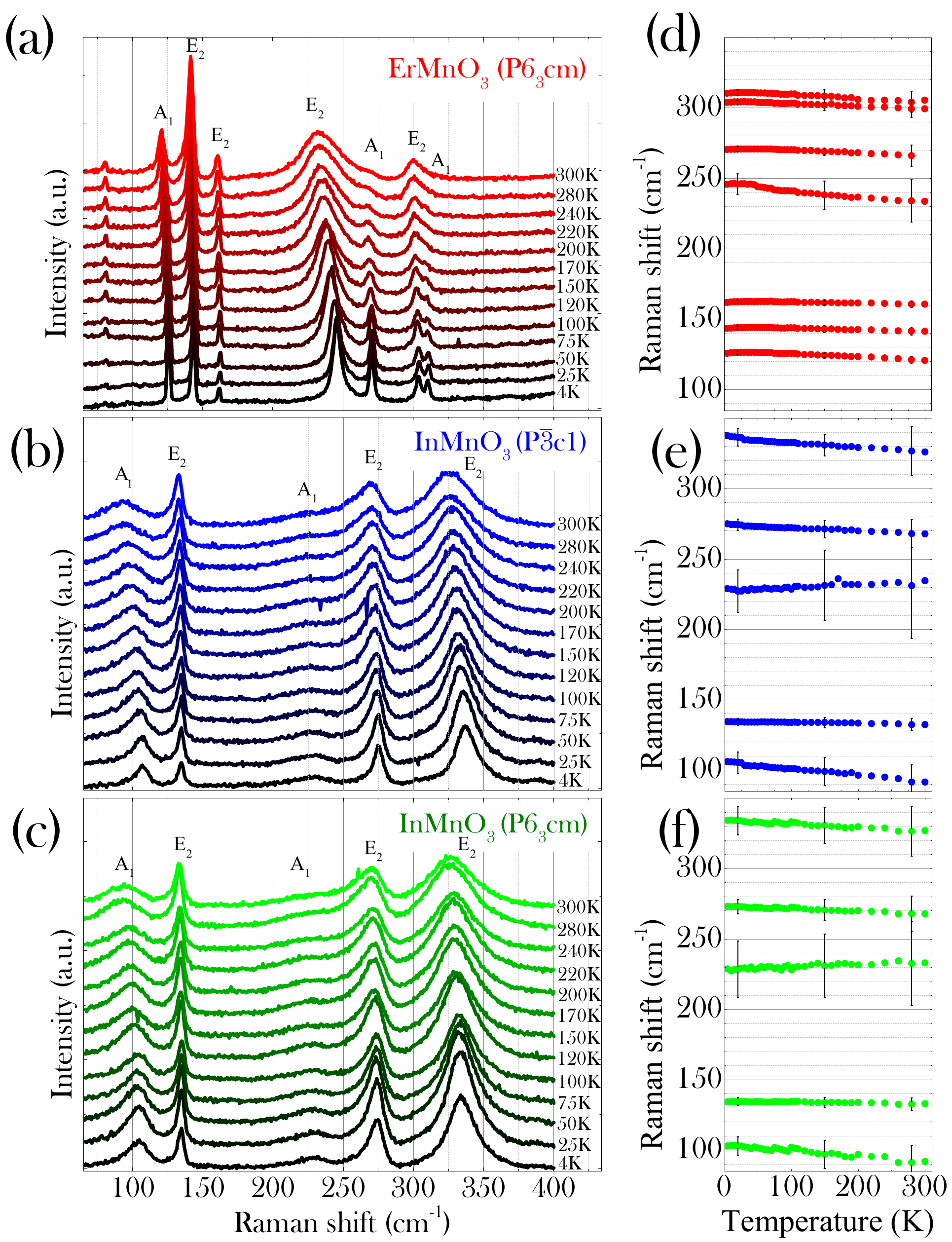}
\caption{Raman spectra of selected temperature from 4K to 300K for three hexagonal manganites samples: (a) ErMnO$_3$, (b) InMnO$_3$ (P$\bar{3}$c1) and (c) InMnO$_3$ (P6$_3$cm). (d-f) Temperature dependence of the position of the modes for the three samples: (d) ErMnO$_3$, (e) InMnO$_3$ (P$\bar{3}$c1) and (f) InMnO3 (P6$_3$cm).}\label{fig_raman}
\end{figure}

In Fig.~\ref{fig_raman}(a-c) we show the detailed temperature dependence of the Raman spectra of all three crystals. For sake of clarity the curves have been shifted vertically proportional to their temperatures. By fitting the curves with Lorentzian functions, we extracted the temperature dependence of the phonon frequencies, shown in Fig.~\ref{fig_raman}(d-f) for (d) the ErMnO$_3$ sample, (e)  the $P\bar{3}c1$ InMnO$_3$ sample and (f) the $P6_3cm$ InMnO$_3$ sample. 

We begin by analyzing the ErMnO$_3$ spectrum, which shows all the A1 and E2 Raman active modes reported previously in the literature \cite{vermetteRamanStudySpin2008} with the expected relative intensities and positions. Moreover, we observe a new small peak below 80 cm$^{-1}$ that was not resolved in the 10K spectrum and increases in intensity with increasing temperature, likely due to the anharmonicity of the potential energy surface. We also observe a general softening of all the modes as the temperature is increased. The frequency of the lowest-frequency A1 mode, which has the strongest Higgs character, reduces by $\sim$10 cm$^{-1}$ between 10 and 300 K, with the higher energy A1 modes reducing in frequency by a similar amount. Since the ferroelectric phase transition in ErMnO$_3$ occurs at $\sim$1200K, the 300 K limit of our experiment corresponds to $\frac{T-T_C}{T_C} = -0.75$, which we see from Figure~\ref{fig:ev_hgs} (b) corresponds to a predicted Landau theory drop in frequency of around 10 cm$^{-1}$, consistent with the experiment. The high $T_C$ of ErMnO$_3$ means that definitive experimental confirmation of Higgs behavior in ErMnO$_3$ would require measurement of the phonon frequencies to higher temperature than is available in our setup. The E1 mode at 250~cm$^{-1}$ shows a particularly strong broadening and redshift; we suggest that this corresponds to a shear mode which we find in our quasi-harmomic calculations to be particularly sensitive to the change in lattice parameters.  

Next we analyze the InMnO$_3$ spectra. Our first observation is that, despite their different ground-state crystal structures, the Raman spectra of the two InMnO$_3$ crystals are almost identical, confirming the similarity in the shapes of their Mexican hat potentials. In both InMnO$_3$ cases, the main Higgs excitation associated with the lowest frequency mode is lower in frequency and softens more rapidly with increasing temperature than in ErMnO$_3$, consistent with the lower Curie temperature of $\sim$500 K. Once again we find a good agreement with the Landau theory prediction, with the calculated drop in frequency between zero and 300 K (corresponding to $\frac{T-T_C}{T_C} = -0.4$) of around 30 cm$^{-1}$ compared with the measured value of $\sim$20 cm$^{-1}$. Interestingly, for InMnO$_3$ the E2 modes show a much weaker temperature dependence than in ErMnO$_3$, with the mode at 135~cm$^{-1}$  largely temperature independent and the mode at 225~cm$^{-1}$ even hardening upon increasing the temperature. We attribute this behaviour to the In-O covalency, which is known to lead to a larger $c$ lattice parameter for InMnO$_3$ compared to other members of the hexagonal manganite series \cite{kumagaiObservationPersistentCentrosymmetricity2012}, and likely also causes markedly different changes in lattice parameters with thermal expansion \cite{bekheetFerroelectricInMnO3Growth2016}.

\section{Summary}
In summary, we have analyzed the role of phonons as Higgs and Goldstone modes at the structural phase transitions in crystalline materials, especially focusing on the case of the hexagonal manganites. We showed that, in materials containing atoms of more than one mass, the static Higgs and Goldstone modes only map uniquely onto single phonon modes at $T_C$, where both the Higgs and  Goldstone frequencies are zero. Below $T_C$, the different masses of the ions cause a softening of several phonon modes with the same symmetry as the static soft-mode distortion. Nevertheless, in both ErMnO$_3$ and InMnO$_3$, our Landau theory analysis identified one primary A1 phonon corresponding to the Higgs mode, and one main B1 phonon corresponding to the Goldstone-like mode. Using Raman spectroscopy, we showed that the lowest A1 modes in both ErMnO$_3$ and InMnO$_3$ indeed have a red shift in the frequency on warming. For InMnO$_3$, in which the temperature range measured is substantial with respect to the Curie temperature, the magnitude of the shift is also substantial, and similar to that predicted by the Landau theory. Therefore we propose that the lowest A1 modes in InMnO$_3$ can be identified as the Higgs modes. The good match between our calculated temperature evolution using Landau and density functional theories, and our Raman measurements suggests that the phase transition in InMnO$_3$ is well described within a standard displacive picture. 

A definitive confirmation of the Higgs mode in ErMnO$_3$, which has a much higher $T_C$, will require Raman measurements to higher temperature. To motivate such measurements, we suggest in addition that ErMnO$_3$ might show intriguing deviations from the behavior that we calculated within Landau theory, since it will likely display similar strong order-disorder behavior to that recently identified in the related YMnO$_3$ \cite{skjaervoUnconventionalContinuousStructural2019}. In an order-disorder transition, the softening of the phonon branches is limited, as observed in inelastic neutron scattering measurements for YMnO$_3$ \cite{guptaSpinphononCouplingHightemperature2015,bansalMomentumresolvedObservationsPhonon2018,bouyanfifHightemperatureLatticedynamicsEvolution2015}  and should be replaced by the emergence of a central peak, which has not yet been identified.

\acknowledgements
We thank M. Bieringer, Department of Chemistry, University of Manitoba, for providing the InMnO$_3$ single crystals and K. Kohn, Waseda University, for providing the ErMnO$_3$ single crystals. We also thank A. Cano and A. Mozzafari  for helpful discussions and M. Miller for graphical assistance. Computing resources were provided by the Euler cluster at ETHZ and CSCS under project IDs eth3 and s889. This work was supported by the Swiss National Science Foundation through the National Center of Competence in Research (NCCR) MARVEL, by the ETH Zurich, and has received funding from the European Research Council (ERC) under the European Union’s Horizon 2020 research and innovation programme grant agreement No 810451. S.M.G. was supported by the Quantum Information Science Enabled Discovery (QuantISED) for High Energy Physics (KA2401032) at LBNL. Work at the Molecular Foundry (S.M.G) was supported by the Office of Science, Office of Basic Energy Sciences, of the U.S. Department of Energy under Contract No. DE-AC02-05CH11231.

\bibliography{references,thesis,Nicola}
\clearpage
\begin{widetext}
\section{Appendix}
\appendix

\begin{table}[h]
\centering
\resizebox{0.15\columnwidth}{!}{
\begin{tabular}{|l|l|}
\toprule
 Frequency [cm$^{-1}$] & Irrep \\
\midrule
             64.8 &                         B1 \\
             86.3 &                         E2 \\
             95.6 &                         A2 \\
            114.4 &                         B2 \\
            128.0 &                         A1 \\
            134.3 &                         E2 \\
            159.7 &                         E1 \\
            167.5 &                         E1 \\
            168.9 &                         E2 \\
            174.3 &                         E2 \\
            176.0 &                         E1 \\
            178.3 &                         B2 \\
            189.2 &                         E2 \\
            193.2 &                         B2 \\
            202.7 &                         A1 \\
            204.0 &                         E1 \\
            204.2 &                         B1 \\
            210.3 &                         E2 \\
            216.1 &                         E1 \\
            230.7 &                         A2 \\
            235.1 &                         A1 \\
            255.8 &                         B2 \\
            272.1 &                         E1 \\
            289.1 &                         A2 \\
            292.0 &                         E2 \\
            303.0 &                         A1 \\
            302.9 &                         B1 \\
            307.8 &                         B2 \\
            340.7 &                         E2 \\
            356.3 &                         E1 \\
            378.9 &                         E1 \\
            380.8 &                         E2 \\
            392.3 &                         B2 \\
            397.2 &                         E1 \\
            404.4 &                         A1 \\
            406.6 &                         E2 \\
            408.9 &                         E1 \\
            410.2 &                         E2 \\
            429.6 &                         A2 \\
            429.9 &                         B1 \\
            445.7 &                         B2 \\
            446.9 &                         A1 \\
            470.8 &                         A1 \\
            494.0 &                         B2 \\
            494.8 &                         A2 \\
            495.6 &                         B1 \\
            501.7 &                         E2 \\
            509.7 &                         E1 \\
            527.5 &                         E1 \\
            533.2 &                         E2 \\
            573.7 &                         A1 \\
            609.3 &                         E1 \\
            611.6 &                         E2 \\
            614.8 &                         E2 \\
            618.9 &                         E1 \\
            645.4 &                         B2 \\
            657.0 &                         A1 \\
            736.5 &                         B2 \\
\bottomrule
\end{tabular}
}
\resizebox{0.1515\columnwidth}{!}{
\begin{tabular}{|l|l|}
\toprule
 Frequency [cm$^-1$] & Irrep \\
\midrule
             76.3 &                         E2 \\
             84.8 &                         A2 \\
            103.9 &                         B2 \\
            107.3 &                         B1 \\
            130.3 &                         A1 \\
            154.3 &                         E2 \\
            155.5 &                         E1 \\
            159.7 &                         B2 \\
            162.1 &                         E1 \\
            163.9 &                         E2 \\
            169.8 &                         E2 \\
            171.0 &                         E1 \\
            206.0 &                         E2 \\
            207.6 &                         E1 \\
            225.0 &                         B2 \\
            245.9 &                         A1 \\
            247.6 &                         E2 \\
            256.2 &                         E1 \\
            263.2 &                         A2 \\
            270.2 &                         B1 \\
            271.5 &                         B2 \\
            278.3 &                         A1 \\
            279.1 &                         B1 \\
            293.4 &                         E1 \\
            300.5 &                         E2 \\
            315.2 &                         A2 \\
            315.9 &                         A1 \\
            341.3 &                         E2 \\
            347.4 &                         B2 \\
            364.0 &                         E1 \\
            385.4 &                         E2 \\
            390.2 &                         E1 \\
            400.7 &                         A2 \\
            409.9 &                         B2 \\
            412.8 &                         E2 \\
            416.4 &                         E1 \\
            424.0 &                         B1 \\
            426.2 &                         A1 \\
            433.1 &                         E1 \\
            451.2 &                         E2 \\
            454.5 &                         E2 \\
            456.5 &                         E1 \\
            458.2 &                         B2 \\
            463.7 &                         A1 \\
            494.7 &                         E1 \\
            495.7 &                         E2 \\
            505.0 &                         A1 \\
            521.2 &                         B1 \\
            526.4 &                         A2 \\
            535.9 &                         B2 \\
            603.1 &                         A1 \\
            632.0 &                         E2 \\
            632.3 &                         E1 \\
            637.4 &                         E2 \\
            637.9 &                         E1 \\
            686.2 &                         B2 \\
            688.9 &                         A1 \\
            771.5 &                         B2 \\
\bottomrule
\end{tabular}
}
\caption{Calculated zero-kelvin DFT phonon frequencies for InMnO$_3$ (left) and ErMnO$_3$ (right).}
\label{Table_PhononFrequencies}
\end{table}
\clearpage

\begin{table}
InMnO$_3$\\
\begin{tabular}{rrrrrrr}
\toprule
  $\omega_{A1}$ &  $\omega_{B1}$ &         c &         d &         e &         f &         g \\
\midrule
       127.8 &         63.4 &  0.9 &   9.2 &  0.2 &  0.5 &  1.5 \\
       127.9 &        204.8 &  0.9 &   1.5 &  0.2 &  0.7 &  0.6 \\
       202.9 &         63.4 &  2.8 &  11.0 &  0.7 &  0.5 &  2.7 \\
       202.9 &        204.8 &  2.8 &   2.0 &  0.7 &  0.7 &  1.4 \\
       236.0 &         63.6 & -6.7 &   2.5 &  1.7 &  0.5 &  0.2 \\
\bottomrule
\end{tabular}

ErMnO$_3$\\
\begin{tabular}{rrrrrrr}
\toprule
 $\omega_{A1}$ &  $\omega_{B1}$ &         c &         d &         e &         f &         g \\
\midrule
       129.6 &        106.1 &  0.8 &   3.3 &  0.0 &  0.1 &  0.1 \\
       129.6 &        267.0 &  0.9 &   3.9 &  0.1 &  0.9 &  0.5 \\
       129.6 &        273.4 &  0.8 &   5.3 &  0.1 &  0.8 &  0.2 \\
       242.3 &        105.9 &  7.8 &   7.4 &  0.8 &  0.1 &  0.9 \\
       242.5 &        267.5 &  8.0 &  10.3 &  0.8 &  0.8 &  2.5 \\
       242.6 &        274.1 &  7.9 &   3.1 &  0.7 &  0.5 &  1.1 \\
       273.6 &        105.4 & -7.7 &   4.6 &  0.4 &  0.2 &  0.3 \\
       273.3 &        267.2 & -7.7 &  17.0 &  0.5 &  0.8 &  0.2 \\
       273.4 &        273.6 & -7.8 &  19.8 &  0.5 &  0.7 &  0.7 \\
       314.5 &        105.6 & -8.9 &  -2.2 &  1.3 &  0.2 &  0.0 \\
       314.5 &        267.6 & -8.9 &  -7.6 &  1.4 &  0.7 &  1.8 \\
       314.5 &        274.1 & -8.9 &  11.6 &  1.4 &  0.5 &  1.0 \\

\bottomrule
\end{tabular}

\caption{Phonon-phonon coupling between A1 and B1 modes in InMnO$_3$ (top) and ErMnO$_3$ (bottom), calculated in this work. Frequencies are given in cm$^{-1}$, units of the coupling constants are meV(amu\text{\AA})$^{-3/2}$ for c and d, and meV/(amu\text{\AA})$^{-2}$} for e, f and g. Slight differences in the frequencies to the calculated values are due to fitting. Only the lowest lying A1 and B1 modes were considered.
\label{appendix:phonon_coupling}
\end{table}
\end{widetext}
\end{document}